\documentclass[aps,prd,amsfonts,nofootinbib,supercriptaddress]{revtex4}
\usepackage[utf8]{inputenc}
\usepackage{amssymb,amsmath}
\usepackage[pdftex]{hyperref}
\usepackage[svgnames]{xcolor}
\definecolor{phthaloblue}{rgb}{0.0, 0.06, 0.54}
\definecolor{bluscuro}{rgb}{0.15, 0.2, .85}
\definecolor{indigo(dye)}{rgb}{0.0, 0.25, 0.42}
\hypersetup{
    colorlinks=true,
    linkcolor=indigo(dye),
    citecolor=indigo(dye),
    filecolor=indigo(dye),
    urlcolor=indigo(dye),
    }
\makeatletter \def\@eqnnum{{\normalsize \normalcolor (\theequation)}}  \makeatother

\begin{document}
\hfill KEK-TH-1966

\title{Trans-Planckian quantum corrections and 
inflationary vacuum fluctuations of non-minimally coupled scalar fields}
\author{Hiroki Matsui,}
\email{hiroki.matsui.c6@tohoku.ac.jp}
\affiliation{Department of Physics, Tohoku University, Sendai, Miyagi 980-8578, Japan}
% e-mail addresses: one for each author, in the same order as the authors

\begin{abstract} \noindent
In the present paper we discuss how
trans-Planckian physics affects inflationary vacuum fluctuations
and primordial density perturbations.
The trans-Planckian problem during inflation 
has been widely discussed in literature, but it is still under debate. 
We reconsider this problem by using
the two-point correlation function
of the non-minimally coupled scalar fields 
and constructing the effective potential with the adiabatic (WKB) regularization or approximation.
First, we clearly show that the cut-off divergence of the quantum fluctuations does not drastically 
change during inflation under reasonable assumptions and the corrections  
can be embedded in standard effective potential.
Thus, the UV effects on the primordial density perturbation
are well translated into the effective potential.
Then, we find out  the modified effective potential from the
inflationary fluctuations and show  
how the trans-Planckian or UV corrections change the potential during inflation.
We clearly show that the new physics strongly affects the inflation potential during inflation
and we obtain a inflationary constraint $\Lambda_{\rm UV} \ll H/g^{1/2}$
where $g$ is the interaction coupling at the UV scale $\Lambda_{\rm UV}$.
\end{abstract}

\maketitle

%%%%%%%%%%%%%%%%%%%%%%%%%%%%%%%%%%%%%%%%%%%%%
\section{Introduction}
%%%%%%%%%%%%%%%%%%%%%%%%%%%%%%%%%%%%%%%%%%%%%
Inflation~\cite{STAROBINSKY198099,Guth:1980zm,Sato:1980yn,Linde:1981mu,Albrecht:1982wi}
is the most standard cosmological paradigm describing the early Universe and
there are accumulating observational evidences which support its existence (see e.g. Ref.~\cite{
Akrami:2018odb}). Famously, the inflation assumes a quasi de Sitter expansion and 
explains for various initial condition problems 
of the standard big bang cosmology elegantly.
However, one of the most attractive aspects of inflation is that the
quantum fluctuations of the scalar fields during inflation 
lead to the primordial density perturbations~\cite{Mukhanov:1981xt,Hawking:1982cz,
Guth:1982ec,Starobinsky:1982ee,Mukhanov:1990me}.
The quantum fluctuations are frozen 
after the wavelength becomes larger than the Hubble radius, and finally become the 
classical density perturbations.
The primordial inhomogeneities generate large-scale structure of the Universe
and provide the highly precise test of the inflation with 
the cosmic microwave background (CMB) anisotropies.

The inflationary trans-Planckian problem~\cite{Niemeyer:2000eh,Brandenberger:2000wr,
Martin:2000xs,Tanaka:2000jw,Starobinsky:2001kn,Starobinsky:2002rp,
Hui:2001ce,Niemeyer:2001qe,Kaloper:2002uj,Kaloper:2002cs,Burgess:2002ub,Brandenberger:2002hs,
Elgaroy:2003gq,Greene:2004np,Greene:2005wk,Danielsson:2002kx,Danielsson:2002mb,Danielsson:2002qh,Danielsson:2005cc,Danielsson:2006gg,
Goldstein:2002fc,Easther:2002xe,Chung:2003wn,Kaloper:2003nv,Burgess:2003hw,Alberghi:2003am,Martin:2003kp,
Meerburg:2010rp,Kundu:2011sg,Groeneboom:2007rf,Ashoorioon:2013eia,Ashoorioon:2014nta,Ashoorioon:2017toq,Broy:2016zik}
suggests that inflationary perturbations provide important clues about trans-Planckian or 
ultraviolet (UV) physics, which has been widely discussed and still under debate. 
In standard paradigm of the inflation we usually assume that the quantum fluctuations 
are valid up to an infinite short length.
But, it is not realistic since new physics is expected to be
below the Planck scale and this simple assumption would be incorrect.
There has been many discussions about how trans-Planckian or UV physics 
affect inflationary perturbations and 
leave a specific imprint on the CMB.
Naively, the inflationary perturbations are not drastically affected by 
the trans-Planckian physics as far as $H \approx \Lambda_{\rm UV}\approx M_{\rm P}=\sqrt{G}$ 
where $H$ is the Hubble constant parameter or unless the Lorentz invariance is broken
(see e.g. Ref.~\cite{Starobinsky:2001kn,Starobinsky:2002rp}). 
However, even if these effects are sufficiently small, it could provide important clues about
the high-energy scale physics.
For instance, taking the initial vacuum as $\alpha$-vacua defined 
at the finite time~\cite{Allen:1985ux,Mottola:1984ar},
the Bogoliubov coefficients ${ \alpha  }_{ k }$ and ${ \beta  }_{ k }$ 
are constrained by the following relation,
\begin{align}
{ \beta  }_{ k }=\frac { i{ e }^{ -2ik{ \eta  }_{ 0 } } }{ 2k{ \eta  }_{ 0 }+i } { \alpha  }_{ k },
\end{align}
where $\eta_{0}$ is the conformal initial time and $k$ is the wave mode.
Note that the Bunch-Davies vacuum is restored in the infinite past
(${ \eta  }_{ 0 }\rightarrow -\infty $, ${ \alpha  }_{ k }=1$ and ${ \beta  }_{ k }=0$).
Initial condition should be imposed when
the wavelength crosses to some fundamental length scale.
Thus, the initial condition could be imposed 
at the $k$-dependent initial time $\eta_{0} 
= -\Lambda_{\rm UV}/Hk$
where $\Lambda_{\rm UV}$ is the cut-off scale.
On the other hand, based on the effective field theory approach,
the inflationary fluctuation can be written as follows~\cite{Kaloper:2002uj,Kaloper:2002cs}:
\begin{align}
\left< { \delta \phi  }^{ 2 } \right> \Bigr|_{k\approx H} = { H }^{ 2 }
+{ c }_{ 1 }{ H }^{ 2 }\left( \frac{{ H }^{ 2 }}{\Lambda_{\rm UV}^{ 2 }}\right)
+{ c }_{ 2 }{ H }^{ 2 }\left( \frac{{ H }^{ 2 }}{\Lambda_{\rm UV}^{ 2 }}\right)^{ 2 }
+{ c }_{ 3 }{ H }^{ 2 }\left( \frac{{ H }^{ 2 }}{\Lambda_{\rm UV}^{ 2 }}\right)^{ 3 }+\cdots 
\end{align}
where the coefficients ${ c }_{ i }$ are determined by the UV physics
and $\mathcal{O}\left( H^2/\Lambda_{\rm UV}^2 \right)$
matches the standard effective field theory approach which 
introduces the higher-dimensional operators and capture the influence of the UV physics.
There is various estimation of the UV corrections of the inflationary fluctuation 
on the CMB power spectrum, and various discussion  
to explore physics at very short distance scales from the CMB measurements.

In this paper we discuss how the UV or 
trans-Planckian physics affects inflationary vacuum fluctuations
and primordial density perturbations.
%%%%%%%%%%%%%%%%%%%%%%%%%%%%%%%%%%%%%%%%%%%%%
\footnote{
The trans-Planckian problems should be discussed 
in the framework of quantum gravity (QG).
However, there is no consistent theory~\cite{DeWitt:2007mi,Smolin:2003rk,Giddings:2011dr}
due to the non-renormalizable or non-unitary properties.
Thus, it is natural to adopt the semiclassical approach to the gravity~\cite{birrell1984quantum,fulling1989aspects,parker2009quantum,
DeWitt:1975ys,Bunch:1979uk,buchbinder1980effective,Shapiro:2008sf}.}
%%%%%%%%%%%%%%%%%%%%%%%%%%%%%%%%%%%%%%%%%%%%%
First, we consider the two-point correlation function $\left< {  \delta \phi  }^{ 2 } \right>$
of the non-minimally coupled scalar fields using the adiabatic (WKB) regularization or approximation
and clearly show that the cut-off divergence of the quantum fluctuations does not drastically 
change during inflation under.
Then, the quantum corrections can be embedded in the effective potential
and therefore, the UV effects on the primordial density perturbation
are well translated into the effective potential.
Then, we construct modified effective potential from the
inflationary fluctuations and show  
how the trans-Planckian or UV corrections change the potential during inflation.
We demonstrate that the new physics drastically changes the inflationary perturbations.

%%%%%%%%%%%%%%%%%%%%%%%%%%%%%%%%%%%%%%%%%%%%%
\section{The UV divergence of quantum fluctuations }
\label{sec:renormalization}
%%%%%%%%%%%%%%%%%%%%%%%%%%%%%%%%%%%%%%%%%%%%%
The quantum fluctuation necessarily causes a problem about renormalization. 
In quantum field theory (QFT), the two-point correlation function 
$\left<{  \delta \phi   }^{ 2 } \right>$  which express the quantum fluctuation
have the UV (quadratic and logarithmic) divergences and
therefore some regularization or renormalization methods are required.
In flat spacetime the divergences of the quantum fluctuation
can be eliminated by the bare parameters of the Lagrangian 
thorough the standard renormalization technique.
However, in curved spacetime~\cite{birrell1984quantum} 
due to the quantum particle creations
the representation of the quantum fluctuation and the renormalization 
have some ambiguity.
In this section, let us revisit the renormalization 
of the quantum fluctuation in de Sitter spacetime
using the adiabatic (WKB) regularization or approximation~\cite{Zeldovich:1971mw,bunch1980adiabatic,
Parker:1974qw,Fulling:1974pu,Fulling:1974zr,birrell1978application,
Anderson:1987yt,Haro:2010zz,Haro:2010mx,Kohri:2017iyl} 
which is a powerful method to remove the divergences.
Consequently, we found out that the inflationary fluctuation or particle creation 
during inflation are sequestered from the UV or trans-Planckian physics.

Through this paper, we consider a spatially flat 
Friedmann-Lemaitre-Robertson-Walker (FLRW) spacetime,
\begin{align}
ds^{2}:=-dt^{2}+a^{2}\left(t\right)\delta_{ij}dx^{i}dx^{j},
\end{align}
where $a\left(t\right)$ is the scale factor and the scalar curvature is give as
$R=6{ \left( { \dot { a }  }/{ a }  \right)  }^{ 2 }+6\left( { \ddot { a }  }/{ a }  \right)  
=6\left({a''}/{a^{3}}\right)$
where $\eta$ is the conformal time and defined by $d\eta=dt/a$.
The scale factor becomes $a\left(t\right) =e^{Ht}$ and 
the scalar curvature is expressed as $R=12H^{2}$ in de Sitter spacetime.

We assume two non-minimal coupled scalar fields 
with the interaction coupling $g$ where inflaton field is $\phi$,
massive scalar field is $S$ with $m_{S} \sim \Lambda_{\rm UV}\gg m_{\phi}$.
The bare Lagrangian is defined by,
\begin{align}
\begin{split}
\mathcal{L}\left[ \phi, S   \right] &:=
\frac { 1 }{ 2 } { g }^{ \mu \nu  }
{ \partial   }_{ \mu  }\phi {\partial  }_{ \nu  }\phi+\frac{1}{2}\left(m^{2}_{\phi}+\xi_{\phi} R\right)\phi^{2}+\frac{\lambda}{4}\phi^{4} \\
&+\frac { 1 }{ 2 } { g }^{ \mu \nu  }{ \partial   }_{ \mu  }S{\partial  }_{ \nu  }S 
+\frac{1}{2}\left(m^{2}_{S}+\xi_{S} R\right)S^{2}
+\frac{g}{2}\phi^{2}S^{2} \label{eq:dddgedg},
\end{split}
\end{align}
where $\xi_{\phi},\xi_{S}$ are the non-minimal curvature couplings.
The Klein-Gordon equations for these two scalar field are given as follows:
\begin{align}
&\Box \phi+m^{2}_{\phi}\phi+\xi_{\phi} R\phi
+\lambda\phi^{3}+{g}S^{2}\phi =0 \label{eq:dsssssdg}, \\
&\Box S+m^{2}_{S}S+\xi_{S} R S+{g}\phi^{2}S =0 \label{eq:dssdssdg},
\end{align}
where 
\begin{align}
\Box =g^{\mu\nu}{ \nabla  }_{ \mu  }{  \nabla   }_{ \nu  }=1/\sqrt { -g } 
{ \partial  }_{ \mu  }\left( \sqrt { -g } { \partial  }^{ \mu  } \right) 
\end{align}
which express generally covariant d'Alembertian operator.
Next, we treat scalar fields $\phi $, $S$ as 
the field operators acting on the ground states
and these scalar fields $\phi $, $S$ are decomposed 
into the classic parts and the quantum parts;
\begin{align}
\phi =\phi \left(\eta,x \right)+\delta \phi \left(\eta ,x \right), \quad
S =S \left(\eta,x \right)+\delta S\left(\eta ,x \right) 
\end{align}
and we assume $\left< 0  \right| { \delta\phi \left(\eta ,x\right) }
\left| 0 \right>=\left< 0  \right| { \delta S \left(\eta ,x\right) }\left| 0 \right>=0$.
Introducing the renormalized parameters and counter-terms to be
\begin{align}
m_{\phi}^{2}=m^{2}_{\phi}\left(\mu\right)+\delta m^{2}_{\phi},\
\xi_{\phi}=\xi_{\phi}\left(\mu\right)+\delta \xi_{\phi}, \
\lambda=\lambda\left(\mu\right)+\delta \lambda, \ 
g=g\left(\mu\right)+\delta g,
\end{align}
the one-loop Klein-Gordon equations of the inflaton field are written as
\begin{align}
&\Box \phi +\left(m^{2}_{\phi}\left(\mu\right)+\delta m^{2}_{\phi}\right)\phi +\left(\xi_{\phi}\left(\mu\right)
+\delta \xi_{\phi}\right)R\phi \nonumber \\ &+3\left(\lambda\left(\mu\right)+\delta \lambda \right)
\left< \delta { \phi  }^{ 2 } \right>\phi +\left(\lambda\left(\mu\right)+\delta \lambda \right)\phi^{3} 
+\left(g\left(\mu\right)+\delta g \right)\left< \delta { S  }^{ 2 } \right>\phi=0, \label{eq:lklkedg} \\
&\left(\Box +m^{2}_{\phi}\left(\mu\right)+\xi_{\phi}\left(\mu\right) R
+3\lambda\left(\mu\right)\phi^{2}+g\left(\mu\right)S^{2}   \right)\delta \phi =0, \nonumber
\end{align}
where the first equation shows the dynamics of the average inflaton field, whereas
the second equation shows the quantum fluctuation of the inflaton field.

The quantum field $\delta \phi$ can be decomposed into each $k$ modes by
\begin{align}
\delta \phi \left( \eta ,x \right) =
\int { { d }^{ 3 }k\left( { a }_{ k }\delta { \phi  }_{ k }\left( \eta ,x \right) 
+{ a }_{ k }^{ \dagger  }\delta { \phi  }_{ k }^{ * }\left( \eta ,x \right)  \right)  }  
\label{eq:ddfkkfledg},
\end{align}
where the creation and annihilation operators of $\delta { \phi  }_{ k }$
are required to satisfy the standard commutation relations
$\bigl[ { a }_{ k },{ a }_{ k' }\bigr] =\bigl[ { a }_{ k }^{ \dagger  },
{ a }_{ k' }^{ \dagger  } \bigr] =0$ and
$\bigl[ { a }_{ k },{ a }_{ k' }^{ \dagger  } \bigr] ={ \delta  }\left( k-k' \right)$.
The in-vacuum state $\left|  0 \right>$ is defined by 
$a_{k}\left|  0 \right>=0 $ and corresponds to the initial conditions 
of the mode functions of $\delta { \phi  }_{ k }$.
The quantum fluctuation $\left<{  \delta \phi   }^{ 2 } \right>$ 
of the inflaton field can be written by
\footnote{
Note that $\Lambda_\textrm{cut-off}$ with the rescaled mode functions $\delta { \chi  }_{ k }\left( \eta  \right)$ corresponds to 
$a\left( \eta  \right)\Lambda_{\rm UV}$.},
\begin{align}
\left< 0  \right| { \delta \phi^{2}  }\left| 0 \right>
=\int_{ 0 }^{ \Lambda_\textrm{UV}  }  { { d }^{ 3 }k{ \left| \delta { \phi  }_{ k }\left( \eta ,x \right)  \right|  }^{ 2 } }
=\frac { 1 }{ 2{ \pi  }^{ 2 }a^{2}\left( \eta  \right)  } \int _{ 0 }^{ \Lambda_\textrm{cut-off}  } { dk { k }^{ 2 }{ \left| \delta { \chi  }_{ k }\left( \eta  \right) \right|  }^{ 2 } }  \label{eq:xkkgdddgedg},
\end{align}
where we introduce the rescaled mode functions 
$\delta { \chi  }_{ k }\left( \eta  \right)$ as ${ \delta \phi  }_{ k }\left( \eta ,x \right)=
e^{ik\cdot x}\delta { \chi  }_{ k }\left( \eta  \right)/{ \left( 2\pi  \right)  }^{ 3/2 }
a\left( \eta  \right)$
From Eq.~(\ref{eq:lklkedg}), 
the Klein-Gordon equation for the quantum rescaled field 
$\delta \chi$ is given by
\begin{align}
{ \delta \chi}''_{ k }\left( \eta  \right)+{ \Omega  }_{ k }^{ 2 }
\left( \eta  \right) { \delta \chi  }_{ k }\left( \eta  \right)=0 \label{eq:dlksdsg},
\end{align}
where
\begin{align}
{ \Omega  }_{ k }^{ 2 }\left( \eta  \right) ={ k }^{ 2 }
+a^{2}\left( \eta  \right) \left( { m }_{\phi}^{ 2 } 
+gS^{2}+3\lambda \phi^{2}+\left( \xi_{\phi} -1/6 \right) R  \right)
={ k }^{ 2 }+a^{2}\left( \eta  \right){ M }^{ 2 }\left( \phi  \right). \nonumber 
\end{align}
Now, we rewrite the rescaled mode function $\delta \chi \left( \eta  \right)$ 
by the Bogoliubov coefficients ${ \alpha  }_{ k }\left( \eta  \right)$,
${ \beta  }_{ k }\left( \eta  \right) $ as
\begin{align}
\delta { { \chi  }_{ k } }\left( \eta  \right) =\frac { 1 }{ \sqrt { 2{ \Omega  }_{ k }
\left( \eta  \right)  }  } \left\{ { \alpha  }_{ k }\left( \eta  \right) 
{ \delta \varphi }_{k}\left( \eta  \right) +{ \beta  }_{ k }\left( \eta  \right)
{ \delta \varphi }_{k}^{*}\left( \eta  \right)  \right\} \label{eq:dldfghdhg},
\end{align}
where ${ \alpha  }_{ k }\left( \eta  \right)$, ${ \beta  }_{ k }\left( \eta  \right) $ 
satisfy the Wronskian condition:
${ \left| { \alpha  }_{ k }\left( \eta  \right)  \right|  }^{ 2 }
-{ \left| { \beta  }_{ k }\left( \eta  \right)  \right|  }^{ 2 }=1$.
The initial conditions for ${ \alpha  }_{ k }\left( \eta_{0}  \right) $,
${ \beta  }_{ k }\left( \eta_{0}  \right) $ 
are equivalent to the choice of the in-vacuum state. 
From Eq.~(\ref{eq:dldfghdhg}) the quantum fluctuation 
$\left<{  \delta \phi   }^{ 2 } \right>$ can be given by~\cite{Ringwald:1987ui}
\begin{align} 
\left<{  \delta \phi   }^{ 2 } \right>
=\frac { 1 }{ 4{ \pi  }^{ 2 }a^{2}\left( \eta  \right)  }\int _{ 0 }^{ \Lambda_\textrm{cut-off}  }dk{ k }^{ 2 }
{ \Omega  }_{ k }^{ -1 }\biggl\{ 1+2{ \left| { \beta  }_{ k } \right|  }^{ 2 }  
+{ \alpha  }_{ k }{ \beta  }_{ k }^{ * }{ \delta \varphi }_{k}^{2}
+{ \alpha  }_{ k }^{ * }{ \beta  }_{ k }{ \delta \varphi^{*}_{k} }^{2} \biggr\}.  
\end{align}
For convenience, we introduce the following quantities $n_{k}={ \left| { \beta  }_{ k } \right|  }^{ 2 }$ 
and $z_{k}={ \alpha  }_{ k }{ \beta  }_{ k }^{ * }{ \delta \varphi }_{k}^{2}$
where $n_{k}={ \left| { \beta  }_{ k }\left( \eta  \right) \right|  }^{ 2 }$ 
can be interpreted as the particle number density created in curved spacetime.
Using $n_{k}$ and $z_{k}$, 
we obtain the following expression of the quantum fluctuation 
of the inflaton field as
\begin{align}\label{eq:adbatigudf}
\left<{  \delta \phi   }^{ 2 } \right>=\left<{  \delta \phi   }^{ 2 } \right>^{({\rm q})}+\left<{  \delta \phi   }^{ 2 } \right>^{({\rm c})} ,
\end{align}
where:
\begin{align*}
\left<{  \delta \phi   }^{ 2 } \right>^{( {\rm q} )}&=
\frac { 1 }{ 4{ \pi  }^{ 2 }a^{2}\left( \eta  \right)  }
\int _{ 0 }^{ \Lambda_\textrm{cut-off}  }{ dk{ k }^{ 2 }{ \Omega  }_{ k }^{ -1 }  },\\
\left<{  \delta \phi   }^{ 2 } \right>^{( {\rm c} )}& =
\frac { 1 }{ 4{ \pi  }^{ 2 }a^{2}\left( \eta  \right)  }
\int _{ 0 }^{ \Lambda_\textrm{cut-off}  }{ dk{ k }^{ 2 }{ \Omega  }_{ k }^{ -1 }
\left\{ 2n_{k}+2{\rm Re}z_{k} \right\}  } ,
\end{align*}
where the first expression 
can be regarded as the classic fluctuations and
expresses particle creations in curved spacetime~\cite{birrell1984quantum} whereas
the second expression obviously diverges and is consistent with 
the flat spacetime.
Therefore, we expect that the first expression is finite and 
insensitivity to the cut-off parameter $\Lambda_\textrm{cut-off}$.

Let us discuss the issues using the adiabatic (WKB) regularization
method~\cite{Zeldovich:1971mw,bunch1980adiabatic,Parker:1974qw,Fulling:1974pu,Fulling:1974zr,birrell1978application,Anderson:1987yt,Haro:2010zz,Haro:2010mx,Kohri:2017iyl}. 
The adiabatic regularization proceed the regularization through subtracting 
$\left<{  \delta \phi   }^{ 2 } \right>^{({\rm q})} $ 
from $\left<{  \delta \phi   }^{ 2 } \right>$ and
calculates the renormalized vacuum fluctuation or the energy density in curved spacetime.
Thus, the inflationary vacuum fluctuation 
can be written as~\cite{Haro:2010zz,Haro:2010mx,Kohri:2017iyl} 
\begin{align}
\begin{split}\label{eq:ohsdedg}
{ \left< { \delta \phi   }^{ 2 } \right>}^{({\rm c})}
&=\int{ \frac { dk }{ k }  }{ P }_{\delta { \phi  } }\left( \eta,  k \right)\\
&={ \left< { \delta \phi   }^{ 2 } \right>}-{ \left< { \delta \phi   }^{ 2 } \right>^{({\rm q})} }
=\frac { 1 }{ 4{ \pi  }^{ 2 }a^{2}\left( \eta  \right)  }
\int _{ 0 }^{ \Lambda_\textrm{cut-off} }{ dk{ k }^{ 2 }{ \Omega  }_{ k }^{ -1 }\left\{2n_{k}+2{\rm Re}z_{k} \right\}  }
\\ &=
\frac { 1 }{ 4{ \pi  }^{ 2 }a^{2}\left( \eta  \right)  } 
\left[\int _{ 0 }^{ \Lambda_\textrm{cut-off}   }{ 2dk{ k }^{ 2 }{ \left| \delta { \chi  }_{ k } \right|  }^{ 2 } }-
\int _{ 0 }^{ \Lambda_\textrm{cut-off}  }{ dk{ k }^{ 2 }{ \Omega  }_{ k }^{ -1 }  } \right] ,
\end{split}
\end{align}
where we must choose an appropriate initial vacuum
and determine the mode function of $\delta \chi \left( \eta  \right)$.
We consider the massive non-minimally coupled case 
(for the details, see Ref.\cite{Haro:2010zz,Haro:2010mx,Kohri:2017iyl})
where the mode function in de Sitter spacetime is given by 
\begin{align}
\delta { \chi  }_{ k }\left( \eta  \right) =\sqrt{\frac{\pi}{4}}\eta^{1/2}\left\{ { \alpha }_{ k }{ H}_{ \nu }^{(2)}\left( k\eta  \right) 
+{ \beta }_{ k }{ H}_{ \nu }^{(1)}\left( k\eta  \right)  \right\} \label{eq:ohghggedg},
\end{align}
where 
${ H}_{ \nu }^{(1, 2 )}\left( k\eta  \right)$ are the Hankel functions
with 
\begin{align}
\nu \equiv \sqrt{\frac{9}{4}-\frac{{ M }^{ 2 }\left( \phi  \right)}{H^{2}}}
\simeq \frac{3}{2}-\frac{{ M }^{ 2 }\left( \phi  \right)}{3H^{2}},
\end{align}
We assume the specific universe from the radiation-dominated stage 
to the de Sitter stage 
and require the matching conditions at $\eta =\eta_{0}$ 
to determine the Bogoliubov coefficients,
\begin{align}
\alpha_{k}=&\frac { 1 }{ 2i } \sqrt { \frac { \pi k{ \eta  }_{ 0 } }{ 2 }  } \biggl( \left( -i+\frac { H }{ 2k }  \right) { H }_{ \nu  }^{ (1)}\left( k{ \eta  }_{ 0 } \right) 
-{ H }_{ \nu  }^{ (1)' }\left( k{ \eta  }_{ 0 } \right)  \biggr) { e }^{ ik/H } \label{eq:ksfdedg}, \\
\beta_{k}=&-\frac { 1 }{ 2i } \sqrt { \frac { \pi k{ \eta  }_{ 0 } }{ 2 }  } \biggl( \left( -i+\frac { H }{ 2k }  \right) { H }_{ \nu  }^{ (2) }\left( k{ \eta  }_{ 0 } \right) 
-{ H }_{ \nu  }^{ (2)' }\left( k{ \eta  }_{ 0 } \right)  \biggr) { e }^{ ik/H } \label{eq:kgdggedg}.
\end{align}
From Eq.~(\ref{eq:ohsdedg}) 
the inflationary fluctuations are given as follows:
\begin{align}
{ \left< { \delta \phi   }^{ 2 } \right>  }^{({\rm c})}
&=\frac { 1 }{ 4{ \pi  }^{ 2 }a^{2}\left( \eta  \right)  }  
\Biggl[ \int _{ 0 }^{ \Lambda_\textrm{cut-off} }{ 2k^{2}{ \left| \delta { \chi  }_{ k } \right|  }^{ 2 }dk } 
-\int _{ \sqrt{2}/\left| \eta \right| }^{ \Lambda_\textrm{cut-off}  }{ dk{ k }^{ 2 }
{ \Omega  }_{ k }^{ -1 }} \Biggr] \nonumber \\
&=\frac { \eta^{2}H^{2} }{ 2{ \pi  }^{ 2 } } \int _{ 0 }^{ H }{ k^{2}{ \left| \delta { \chi  }_{ k } \right|  }^{ 2 }dk }  
+\frac { \eta^{2}H^{2} }{ 2{ \pi  }^{ 2 }} \int _{ H }^{ \sqrt{2}/\left| \eta \right| }{ k^{2}{ \left| \delta { \chi  }_{ k } \right|  }^{ 2 }dk }  
\label{eq:kdddegedg}.
\end{align}
where we introduce the mode cut-off $k> \sqrt{2-{ M }^{ 2 }\left( \phi  \right)/H^{2}}/\left|\eta \right| 
\simeq \sqrt{2}/\left|\eta \right|\simeq \sqrt{2}aH$.
In the limit ${ \Lambda_\textrm{cut-off} \rightarrow \infty }$
the divergence parts exactly cancel~\cite{Haro:2010zz,Haro:2010mx,Kohri:2017iyl},
\begin{align}\label{eq:dffsedg}
\lim _{ \Lambda_\textrm{cut-off} \rightarrow \infty }\frac { 1 }{ 4{ \pi  }^{ 2 }a^{2}\left( \eta  \right)  }  
\Biggl[ \int _{ \sqrt{2}/\left| \eta \right|}^{ \Lambda_\textrm{cut-off}  }{ 2k^{2}
{ \left| \delta { \chi  }_{ k } \right|  }^{ 2 }dk } 
-\int _{ \sqrt{2}/\left| \eta \right| }^{ \Lambda_\textrm{cut-off} }
{ dk{ k }^{ 2 }{ \Omega  }_{ k }^{ -1 }} \Biggr],
\end{align}
Beyond the mode cut-off $k> \sqrt{2}aH$,
the mode function does not drastically change 
against the evolution of the universe.
In this sense the mode cut-off can be recognized as the UV cut-off of the 
inflationary perturbations and therefore the influences of 
the UV or trans-Planckian physics are sequestered as long as the Hubble parameter 
is much smaller than the cut-off scale, $H \ll  \Lambda_{\rm UV}$.

By using the formula of the Hankel functions 
\begin{align}
{ H }_{ \nu  }^{ (1,2)' }\left( k{ \eta  }_{ 0 } \right)=
{ H }_{ \nu-1  }^{ (1,2) }\left( k{ \eta  }_{ 0 } \right)-\frac{\nu}{k{ \eta  }_{ 0 }}{ H }_{ \nu  }^{ (1,2) }\left( k{ \eta  }_{ 0 } \right),
\end{align}
and the Bessel function of 
the first kind defined by $J_{\nu}=( { H }_{ \nu  }^{ (1) }+{ H }_{ \nu  }^{ (2) } )/2$,
we can obtain the expression 
\begin{align}
\left| { \alpha  }_{ k }-{ \beta  }_{ k } \right| =\sqrt { \frac { \pi k }{ 2H }  } \left| { J }_{ \nu -1 }\left( k{ \eta  }_{ 0 } \right) +\left( i-\frac { H }{ 2k } +\frac { \nu H }{ k }  \right) { J }_{ \nu  }\left( k{ \eta  }_{ 0 } \right)  \right| .
\end{align}
For small $k$ modes, 
the the Bessel function and the Hankel function asymptotically behave as
\begin{align}
J_{\nu}\left(k\eta_{0}\right)&\simeq 
\frac { 1 }{ \Gamma \left( \nu +1 \right)  }
 { \left( \frac { k{ \eta  }_{ 0 } }{ 2 }  \right)  }^{ \nu  },\\
{ H }_{ \nu }^{ (2) }\left( k{ \eta  }_{ 0 } \right)&
\simeq -{ H }_{ \nu }^{ (1) }\left( k{ \eta  }_{ 0 } \right)
\simeq \frac { i }{ \pi  } \Gamma \left( \nu  \right) 
{ \left( \frac { k{ \eta  }_{ 0 } }{ 2 }  \right)  }^{ -\nu  }.
\end{align}
Thus, we can obtain the following expression of the mode function,
\begin{align}
{ \left| \delta { \chi  }_{ k } \right|  }^{ 2 } &\simeq
\frac { \pi  }{ 4 } \left| \eta  \right| { \left| { \alpha  }_{ k }-{ \beta  }_{ k } \right|  }^{ 2 }{ { \left| { H }_{ \nu }^{ (2) }\left( k{ \eta  }\right) \right|  }^{ 2 } }
\simeq
\frac{2}{9k}\left(H\left| \eta \right|\right)^{1-2\nu}\quad\ \left( 0\le  k \le  H \right)\label{eq:kddffsedg}.
\end{align}
For large $k$ modes, we approximate 
the Bogoliubov coefficients to be $\alpha_{k} \simeq1$ and $\beta_{k} \simeq 0$
and evaluate the mode function as
\begin{align}
\delta { \chi  }_{ k }\left( \eta  \right) 
\simeq \sqrt{\frac{\pi}{4}}\eta^{1/2}{ H}_{ \nu }^{(2)}\left( k\eta  \right) .
\end{align}
Thus, we can get the following expression 
\begin{align}
{ \left| \delta { \chi  }_{ k } \right|  }^{ 2 } \simeq \frac{\left| \eta \right|}{16}
\left(\frac{k\left| \eta \right|}{2}\right)^{-2\nu}\quad  \left( H\le  k \le \sqrt { 2 } /\left| \eta  \right|  \right).\label{eq:kdyydg}
\end{align}
From Eq.~(\ref{eq:kddffsedg}) and Eq.~(\ref{eq:kdyydg}),
the vacuum fluctuations are written as
\begin{align}
\begin{split}
{ \left< { \delta \phi   }^{ 2 } \right>  }^{({\rm c})}
&\simeq  \frac { \left(H\left| \eta \right|\right)^{3-2\nu} }{ 9{ \pi  }^{ 2 } } \int _{ 0 }^{ H }{ k dk }  +\frac { H^{2}\left| \eta \right|^{3-2\nu} }{ 4{ \pi  }^{ 2 }\cdot2^{3-2\nu}} \int _{ H }^{ \sqrt{2}/\left| \eta \right| }{ k^{2-2\nu}dk }\\
&\simeq \frac { H^{2} }{ 18{ \pi  }^{ 2 } }e^{-\frac{2{ M }^{ 2 }\left( \phi  \right)t}{3H}}
+\frac { 3H^{4} }{ 8{ \pi  }^{ 2 }{ M }^{ 2 }\left( \phi  \right)} 
\left(1-e^{-\frac{2{ M }^{ 2 }\left( \phi  \right)t}{3H}}\right)
\label{eq:kfhshggdg}.
\end{split}
\end{align}
For late cosmic-time ($N_{\rm tot}=Ht\gg H^{2}/{ M }^{ 2 }\left( \phi  \right)$),
the inflationary fluctuations are approximately written as
%%%%%%%%%%%%%%%%%%%%%%%%%%%%%%%%%%%%%%%%%%%%%
\footnote{
The power spectrum on super-horizon scale 
($\left| k\eta \right|  \ll1$) can be approximately written by~\cite{Riotto:2002yw}
\begin{align}
{ P }_{\delta { \phi  } }\left(\eta,k\right)={ \left( \frac { H }{ 2\pi  }  \right)  }^{ 2 }{ \left( \frac { k }{ aH  }  \right)  }^{ 3-2\nu }
\end{align}
which has a tiny $k$-dependence, i.e scale invariance.}
%%%%%%%%%%%%%%%%%%%%%%%%%%%%%%%%%%%%%%%%%%%%%
\begin{align}
{ \left< { \delta \phi   }^{ 2 } \right>  }^{({\rm c})}
\simeq \frac { 3H^{4} }{ 8{ \pi  }^{ 2 }{ M }^{ 2 }\left( \phi  \right)} , \quad\ \left( { M }\left( \phi  \right)\ll H \right)
\label{eq:klkddussg}.
\end{align}
which is consistent with the well-known results using stochastic approach
(see e.g. Ref.~\cite{Finelli:2008zg,Finelli:2010sh}). 
Therefore, the UV divergences are sequestered and 
the influence of the trans-Planckian physics is negligible.
Precisely, however, ${ \left< { \delta \phi   }^{ 2 } \right>  }^{({\rm c})}$ 
is mode dependent and might leave a tiny trans-Planckian 
imprint on the CMB at the short distance.
Furthermore, the inflaton mass should be smaller than
the Hubble scale if not the inflationary fluctuations are strongly 
suppressed~\cite{Mottola:1984ar} and the effective mass is formally given by the 
effective potential.
From Eq.~(\ref{eq:adbatigudf}) the UV divergences 
are essentially sequestered and the quantum corrections
are embedded in the effective potential.
The quantum effects of the UV physics only changes the 
effective potential through the radiative corrections, but they often modify
the slow roll parameter~\cite{Boyanovsky:2005sh,Sloth:2006az}.
In next section we construct modified effective potential from the
inflationary fluctuations and discuss how the UV quantum corrections 
affect the potential during inflation.

%%%%%%%%%%%%%%%%%%%%%%%%%%%%%%%%%%%%%%%%%%%%%
\section{The modified effective potential during inflation}
\label{sec:modified}
%%%%%%%%%%%%%%%%%%%%%%%%%%%%%%%%%%%%%%%%%%%%%
The effective potential receives the backreaction of the inflationary perturbations 
and UV radiative corrections.
In this section we consider the modified effective potential during inflation.
The inflationary expansion is determined by the effective potential
and the classical slow-roll motion of the inflaton field.
The effective potential is constructed by quantum loop corrections
and corresponds to the effective vacuum energy density $\rho_{\rm eff}$
or pressure $P_{\rm eff}$ derived from the effective energy-momentum tensor,
\begin{align}
\rho_{\rm eff}=\frac{1}{2}{ \dot { \phi  }  }^{ 2 }+\frac { 1 }{ 2 } \nabla { \phi  }^{ 2 }
+V_{\rm eff}\left( \phi  \right), \quad 
P_{\rm eff}=\frac{1}{2}{ \dot { \phi  }  }^{ 2 }-\frac { 1 }{ 2 } \nabla { \phi  }^{ 2 }
-V_{\rm eff}\left( \phi  \right).
\end{align}
In the effective Friedman equations the Hubble parameter 
in the slow-roll approximation can be written as 
\begin{equation}
3H^2 \simeq\frac{V_{\rm eff}(\phi)}{M_{\rm p}^2},\qquad
3H\dot\phi\simeq -V'_{\rm eff}(\phi),
\end{equation}
The slow-roll parameters $\epsilon$, $\eta$ related with 
the observed primordial scalar or tensor perturbations can be given as
\begin{equation}
\epsilon \equiv \frac{1}{2M_{\rm p}^2}\frac{\dot\phi^2}{H^2} \ \Longrightarrow \
\epsilon\simeq \frac{4}{3}\frac{M_{\rm eff}^2\left( \phi  \right) }{H^2}\ll 1~,
\end{equation}
where we define $M_{\rm eff}^2(\phi)=V''_{\rm eff}(\phi)$
and the slow-roll approximation require $M_{\rm eff}(\phi)\ll H$.

Let us consider how to treat the quantum fluctuations during inflation using 
Eq.~(\ref{eq:lklkedg}) and Eq.~(\ref{eq:adbatigudf}) again.
We can regularize the UV divergences of the quantum fluctuations 
$\left<{  \delta \phi   }^{ 2 } \right>^{({\rm q})}$ using the dimensional regularization,
\begin{align}
\left<{  \delta \phi   }^{ 2 } \right>^{({\rm q})}
=\frac { { M }^{ 2 }\left( \phi  \right)  }{ 16{ \pi  }^{ 2 } } 
\left[ \ln { \left( \frac { { M }^{ 2 }\left( \phi  \right)  }{ { \mu  }^{ 2 } }  \right)  } 
-N_{\epsilon}-\frac { 3 }{ 2 }   \right]  \label{eq:rehldg},
\end{align}
where:
\begin{align}
{ M }^{ 2 }\left( \phi  \right) ={ m }^{ 2 }_{\phi} +g\,S^{2}
+3\lambda\,\phi^{2}+\left( 12\,\xi_{\phi} -2\right)H^2,
\end{align}
where $N_{\epsilon}=-1/{ \epsilon  } -\log { 4\pi }-\gamma$,
$\gamma$ is the Euler-Mascheroni constant and 
$\mu$ is the renormalization parameter.
From Eq.~(\ref{eq:rehldg})
the quantum fluctuation during inflation can be written as
\begin{align}
\left<{  \delta \phi   }^{ 2 } \right>
=\left<{  \delta \phi   }^{ 2 } \right>^{({\rm q})} 
+ \left<{  \delta \phi   }^{ 2 } \right>^{({\rm c})}
= \frac { { M }^{ 2 }\left( \phi  \right)  }{ 16{ \pi  }^{ 2 } } 
\left[ \ln { \left( \frac { { M }^{ 2 }\left( \phi  \right)  }{ { \mu  }^{ 2 } }  \right)  } 
-N_{\epsilon}-\frac { 3 }{ 2 }   \right]
+ \left<{  \delta \phi   }^{ 2 } \right>^{({\rm c})}. \label{eq:xvxcldg}
\end{align}
which is almost same as flat spacetime except $\left<{  \delta \phi   }^{ 2 } \right>^{({\rm c})}$.
From the above relation we can get
the one-loop effective potential during inflation as follows~\cite{Kohri:2017iyl}:
\begin{align}\label{eq:tutusssdg}
\begin{split}
V_{\rm eff}\left( \phi  \right) &=\ \frac{1}{2}m^{2}_{\phi} \phi^{2}
+6\xi_{\phi}H^2 \phi^{2}
+\frac{\lambda}{4}\phi^{4} +\frac{3\lambda}{2}
\left<{  \delta \phi   }^{ 2 } \right>^{({\rm c})}\phi^{2}
+\frac { { M }^{ 4 }\left( \phi  \right)  }
{ 64{ \pi  }^{ 2 } } \left[ \ln { \left( \frac { {M}^{ 2 }
\left( \phi  \right)  }{ { \mu  }^{ 2 } }  \right)  } -\frac { 3 }{ 2 }    \right] \\
&+\mathcal{O}\left( \left\{\left<{  \delta \phi   }^{ 2 } \right>^{({\rm c})}\right\}^2\right),
\end{split}
\end{align}
where the divergences can be eliminated by the renormalization parameters, 
and the radiative corrections express quantum loop effects.
The modified effective potential has two additional terms, non-minimal coupling terms and 
the back-reaction terms from the inflationary fluctuations
which usually break the slow-roll conditions.
Therefore, we must impose the condition on these couplings,
\begin{equation}\label{eq:condthermal}
\frac{16\,\xi_{\phi}H^2 }{H^2},\
\frac{\lambda\left<{  \delta \phi   }^{ 2 } \right>^{({\rm c})}}{H^2}
\ll 1 \ \Longrightarrow \
\xi_{\phi},\, \lambda \ll 1,
\end{equation}
where we assume the inflationary fluctuations are as large as the expression of 
Eq.~(\ref{eq:klkddussg}).

Next, let us consider the quantum corrections of the massive scalar field $S$
where ${ m }_{S} \gg H$ and construct the effective potential during inflation.
We obtain one-loop effective potential during inflation by using $\left<{  \delta \phi  }^{ 2 } \right>$
and $\left<{  \delta S  }^{ 2 } \right>$,
\begin{align}
V_{\rm eff}\left( \phi  \right)& =\ \frac{1}{2}m^{2}_{\phi}\phi^{2}
+6\xi_{\phi}H^2 \phi^{2}+\frac{\lambda}{4}\phi^{4}+\frac{3\lambda}{2}
\left<{  \delta \phi   }^{ 2 } \right>^{({\rm c})}\phi^{2}+\mathcal{O}\left( \left\{\left<{  \delta \phi   }^{ 2 } \right>^{({\rm c})}\right\}^2\right) \nonumber  \\
&+\frac { { M }^{ 4 }\left( \phi  \right)  }{ 64{ \pi  }^{ 2 } } \left[ \ln { \left( \frac { {M}^{ 2 }\left( \phi  \right)  }{ { \mu  }^{ 2 } }  \right)  } 
-\frac { 3 }{ 2 }    \right] +\frac{g}{2}\phi^{2}S^{2} +\frac { { M }^{ 4 }\left( S \right)  }{ 64{ \pi  }^{ 2 } } \left[ \ln { \left( \frac { { M }^{ 2 }\left( S \right)  }{ { \mu  }^{ 2 } }  \right)  } 
-\frac { 3 }{ 2 }   \right] +\cdots, \label{eq:tutssdg}
\end{align}
where $\left<{  \delta S }^{ 2 } \right>^{({\rm c})}$ 
is sufficiently suppressed and we define 
\begin{align}
{ M }^{ 2 }\left( S \right)={ m }^{ 2 }_{S} +g\phi^{2}
+\left( 12\,\xi_{S} -2 \right) H^2, \nonumber 
\end{align}
By using Eq.~(\ref{eq:tutssdg}), 
we can read off the $\mu$ dependence of these couplings
and the one-loop $\beta$ function of ${ m }^{ 2 }_{\phi}$ can be given as follows:
\begin{align}
\beta_{ m^{2}_{\phi} }= {\mu} \frac { \partial  }{ \partial { \mu  }  }{ { m }^{ 2 }_{\phi} }  =\frac { 6{ \lambda  }m^{2}_{\phi}
+2gm_{S}^{2} }{ { \left( 4\pi  \right)  }^{ 2 } }
+\cdots.
\end{align}
where the renormalization scale $\mu$ express various phenomenological scale;
\begin{equation}
\mu \approx \phi,\, S,\, H
\end{equation}
Recalling that large field inflation takes the Planck field value of the inflaton and 
the inflaton field is larger than the Hubble scale, $\phi \sim M_{\rm P} \gg H$.
Thus, the fine-tuning problem of the inflaton potential arises unless $\mu \sim \phi \ll m_{S}$.
At least the slow-roll condition requires $ { m }_{\phi}\ll H $
and therefore, 
the inflationary fluctuation or the primordial CMB perturbation are highly
depend on the UV physics.
In this sence, for $g^{1/2}{ m }_{S}\gg H $,
the slow-roll condition of the inflation violates or
the inflationary fluctuation breaks 
the scale invariance of the spectrum of CMB perturbations
(see e.g. Ref.\cite{Linde:1982uu,Starobinsky:1982ee,Vilenkin:1983xp}).
Therefore, we can simply impose a constraint on the UV physics as follows: 
\begin{align}\label{eq:condsclar}
{ m }_{S} \ll H/g^{1/2}.
\end{align}
Then, we also get an upper bound 
of the interaction coupling to be $g \ll \mathcal{O}(10^{-10})$
if we take the Planck mass ${ m }_{S}\sim M_{\rm P}$
and the current upper bound of the Hubble parameter~\cite{Ade:2015lrj,Ade:2015tva}.

From here, let us consider the influences of the quantum corrections in more detail.
The inflationary effective mass ${ M }_{\rm eff}\left( \phi  \right)$ including 
the UV radiative corrections is 
\begin{align}
{ M }^{ 2 }_{\rm eff}\left( \phi  \right) &= { m }^{ 2 }_{\phi} +gS^{2}
+3\lambda \phi^{2}  
+\left( 12\,\xi_{\phi} -2 \right)H^2+ \frac { 3\lambda{ M }^{ 2 }\left( \phi  \right)  }{ 16{ \pi  }^{ 2 } }
\ln { \left( \frac { {M}^{ 2 }\left( \phi  \right)  }{ { m }^{ 2 }_{\phi} }  \right)  } 
\nonumber \\ &
+ \frac { g{ M }^{ 2 }\left( S \right)  }{ 16{ \pi  }^{ 2 } }
\ln { \left( \frac { {M}^{ 2 }\left( S  \right)  }{ { m }_{S}^2 }  \right)  } 
+{3\lambda}\left<{  \delta \phi   }^{ 2 } \right>^{({\rm c})}\cdots   \label{eq:ksjdjrujfdssg}.
\end{align}
where we set the renormalization parameter $\mu$ adequately
and the effective mass ${ M }_{\rm eff}\left( \phi  \right)$ takes the bare 
value ${ M }_{\rm eff}\left( \phi  \right)={ m }^{ 2 }_{\phi} $
in the limit $\phi,S,H \rightarrow 0$.
If we assume $g^{1/2}{ M }\left( S \right) \gg H$
the inflationary effective mass completely breaks the slow-roll conditions when 
\begin{align}
{ m }^{ 2 }_{S} \ll g\phi^{2}
+\left( 12\,\xi_{S} -2 \right) H^2, 
\end{align}
which significantly changes the radiative correction terms. Thus, 
even if the inflaton satisfy the slow-roll conditions at the beginning 
of the inflation, at late time the effective mass glows and breaks the conditions.
For the large field inflation $\phi \sim M_{\rm P}$ or the non-minimal coupling case
the slow-roll condition require the decoupling of the inflaton sector and 
the massive scalar field and we can get the following conditions,
\begin{align}\label{eq:conduv}
g \ll \frac{H^2}{{ m }^{ 2 }_{S} +g\phi^{2}
+\left( 12\,\xi_{S} -2 \right) H^2}
\quad {\rm or}\quad { m }^{ 2 }_{S} \gg g\phi^{2}
+\left( 12\,\xi_{S} -2 \right) H^2
\end{align}
From this viewpoint, the inflaton sector should be completely decoupled with any high energy physics.
Thus, the Starobinsky inflation~\cite{STAROBINSKY198099} 
or the Higgs inflation~\cite{Bezrukov:2007ep,Barvinsky:2008ia} are very attractive
since they do not require new physics
%%%%%%%%%%%%%%%%%%%%%%%%%%%%%%%%%%%%%%%%%%%%%
\footnote{
There are well-known strong correspondence~\cite{Shapiro:2008sf,Salvio:2015kka,Calmet:2016fsr} between the Starobinsky inflation ($a_{1} R^{2}$) 
and the Higgs inflation ($\xi_{H}H^{\dagger}H$).}
%%%%%%%%%%%%%%%%%%%%%%%%%%%%%%%%%%%%%%%%%%%%%
in comparison with any other inflation models~\cite{Martin:2013tda,Martin:2013nzq,
Martin:2013gra,Ijjas:2013vea} and furthermore, they matched with the current
constraints of the CMB observations.
The UV corrections of 
the massive scalar field $S$ drastically changes the effective potential 
of the inflaton and sometimes breaks the slow-roll conditions during inflation. 
These issues can be interpreted by using Eq.~(\ref{eq:klkddussg})
and we can get the following expression of the inflationary fluctuation
\begin{align}
\left< {  \delta \phi  }^{ 2 } \right>^{({\rm c})}_{\rm eff}\simeq
\frac{3{ H }^{ 4}}{8{ \pi  }^{ 2 }\bigl( { m }^{ 2 }_{\phi}+ g \Lambda_{\rm UV}^{2} \bigr)}, 
\quad\ \left( { m }_{\phi}+ g^{1/2}\Lambda_{\rm UV} \ll H \right)
\label{eq:klkdslfssg},
\end{align}
where we take $m_{S} \sim \Lambda_{\rm UV}$ and 
include quantum backreactions in Eq.~(\ref{eq:klkddussg}) by hand.
Thus, the quantum effects of the UV or trans-Planckian physics affects 
the primordial density perturbation significantly.
As previously seen in Section~\ref{sec:renormalization} 
we can not easily read the the quantum effects 
of the UV or trans-Planckian physics in the inflationary perturbation
since the quantum fluctuation of the flat spacetime and the curved spacetime 
is indistinguishable in the UV region. 
Constructing the effective potential, we can systemically 
read the trans-Planckian effects on the primordial density perturbation.

%%%%%%%%%%%%%%%%%%%%%%%%%%%%%%%%%%%
%%%%%%%%%%%%%%%%%%%%%%%%%%%%%%%%%%%
\section{Conclusion and Discussion}
\label{sec:relaxation}
%%%%%%%%%%%%%%%%%%%%%%%%%%%%%%%%%%%
%%%%%%%%%%%%%%%%%%%%%%%%%%%%%%%%%%%
In the present paper we discuss how
trans-Planckian physics affects inflationary vacuum fluctuations
and primordial density perturbations.
Here, we consider the two-point correlation function $\left< {  \delta \phi  }^{ 2 } \right>$
of the non-minimally coupled scalar fields and constructs effective potential.
We have clearly shown that the cut-off divergence of the quantum fluctuations does not drastically 
change during inflation and the quantum corrections 
can be systemically embedded in standard effective potential.
We have constructed the modified effective potential from the inflationary fluctuations
using the adiabatic (WKB) regularization or approximation.
The different points compared with the standard effective potential in flat spacetime 
are that Eq.~(\ref{eq:tutusssdg}) and Eq.~(\ref{eq:tutssdg})
includes the backreaction terms of the inflationary fluctuations, non-minimal coupling 
of the gravity and the UV corrections of the massive scalar field.
These terms generally breaks the slow-roll conditions and significantly 
affects the primordial density perturbation.
We have obtained various conditions to succeed the slow-roll inflation 
like Eq.~(\ref{eq:condthermal}),
Eq.~(\ref{eq:condsclar}) and Eq.~(\ref{eq:conduv})
and got the conjecture $\Lambda_{\rm UV} \ll H/g^{1/2}$
where $g$ is the interaction coupling at the new physics scale $\Lambda_{\rm UV}$.
The inflaton sector should be completely decoupled with the UV sector
and, Starobinsky inflation or Higgs inflation are very attractive
from this viewpoint.

\medskip
%%%%%%%%%%%%%%%%%%%%%%%%%%%%
%%%%%%%%%%%%%%%%%%%%%%%%%%%%
\noindent {\it Acknowledgments}: I would like to thank 
Kazunori Kohri for numerous helpful discussions and collaboration.
%%%%%%%%%%%%%%%%%%%%%%%%%%%%
%%%%%%%%%%%%%%%%%%%%%%%%%%%%

%%%%%%%%%%%%%%%%%%%%%%%%%%%%
\nocite{}
\bibliography{Reference2}

\providecommand{\href}[2]{#2}\begingroup\raggedright\begin{thebibliography}{10}

\bibitem{STAROBINSKY198099}
A.~Starobinsky, \emph{A new type of isotropic cosmological models without
  singularity},
  \href{http://dx.doi.org/https://doi.org/10.1016/0370-2693(80)90670-X}{\emph{Physics
  Letters B} {\bf 91} (1980) 99 -- 102}.

\bibitem{Guth:1980zm}
A.~H. Guth, \emph{{The Inflationary Universe: A Possible Solution to the
  Horizon and Flatness Problems}},
  \href{http://dx.doi.org/10.1103/PhysRevD.23.347}{\emph{Phys. Rev.} {\bf D23}
  (1981) 347--356}.

\bibitem{Sato:1980yn}
K.~Sato, \emph{{First Order Phase Transition of a Vacuum and Expansion of the
  Universe}}, {\emph{Mon. Not. Roy. Astron. Soc.} {\bf 195} (1981) 467--479}.

\bibitem{Linde:1981mu}
A.~D. Linde, \emph{{A New Inflationary Universe Scenario: A Possible Solution
  of the Horizon, Flatness, Homogeneity, Isotropy and Primordial Monopole
  Problems}}, \href{http://dx.doi.org/10.1016/0370-2693(82)91219-9}{\emph{Phys.
  Lett.} {\bf 108B} (1982) 389--393}.

\bibitem{Albrecht:1982wi}
A.~Albrecht and P.~J. Steinhardt, \emph{{Cosmology for Grand Unified Theories
  with Radiatively Induced Symmetry Breaking}},
  \href{http://dx.doi.org/10.1103/PhysRevLett.48.1220}{\emph{Phys. Rev. Lett.}
  {\bf 48} (1982) 1220--1223}.

\bibitem{Akrami:2018odb}
{\scshape Planck} collaboration, Y.~Akrami et~al., \emph{{Planck 2018 results.
  X. Constraints on inflation}},  \href{http://arxiv.org/abs/1807.06211}{{\tt
  1807.06211}}.

\bibitem{Mukhanov:1981xt}
V.~F. Mukhanov and G.~V. Chibisov, \emph{{Quantum Fluctuations and a
  Nonsingular Universe}}, {\emph{JETP Lett.} {\bf 33} (1981) 532--535}.

\bibitem{Hawking:1982cz}
S.~W. Hawking, \emph{{The Development of Irregularities in a Single Bubble
  Inflationary Universe}},
  \href{http://dx.doi.org/10.1016/0370-2693(82)90373-2}{\emph{Phys. Lett.} {\bf
  115B} (1982) 295}.

\bibitem{Guth:1982ec}
A.~H. Guth and S.~Y. Pi, \emph{{Fluctuations in the New Inflationary
  Universe}}, \href{http://dx.doi.org/10.1103/PhysRevLett.49.1110}{\emph{Phys.
  Rev. Lett.} {\bf 49} (1982) 1110--1113}.

\bibitem{Starobinsky:1982ee}
A.~A. Starobinsky, \emph{{Dynamics of Phase Transition in the New Inflationary
  Universe Scenario and Generation of Perturbations}},
  \href{http://dx.doi.org/10.1016/0370-2693(82)90541-X}{\emph{Phys. Lett.} {\bf
  117B} (1982) 175--178}.

\bibitem{Mukhanov:1990me}
V.~F. Mukhanov, H.~A. Feldman and R.~H. Brandenberger, \emph{{Theory of
  cosmological perturbations. Part 1. Classical perturbations. Part 2. Quantum
  theory of perturbations. Part 3. Extensions}},
  \href{http://dx.doi.org/10.1016/0370-1573(92)90044-Z}{\emph{Phys. Rept.} {\bf
  215} (1992) 203--333}.

\bibitem{Niemeyer:2000eh}
J.~C. Niemeyer, \emph{{Inflation with a Planck scale frequency cutoff}},
  \href{http://dx.doi.org/10.1103/PhysRevD.63.123502}{\emph{Phys. Rev.} {\bf
  D63} (2001) 123502}, [\href{http://arxiv.org/abs/astro-ph/0005533}{{\tt
  astro-ph/0005533}}].

\bibitem{Brandenberger:2000wr}
R.~H. Brandenberger and J.~Martin, \emph{{The Robustness of inflation to
  changes in superPlanck scale physics}},
  \href{http://dx.doi.org/10.1142/S0217732301004170}{\emph{Mod. Phys. Lett.}
  {\bf A16} (2001) 999--1006},
  [\href{http://arxiv.org/abs/astro-ph/0005432}{{\tt astro-ph/0005432}}].

\bibitem{Martin:2000xs}
J.~Martin and R.~H. Brandenberger, \emph{{The TransPlanckian problem of
  inflationary cosmology}},
  \href{http://dx.doi.org/10.1103/PhysRevD.63.123501}{\emph{Phys. Rev.} {\bf
  D63} (2001) 123501}, [\href{http://arxiv.org/abs/hep-th/0005209}{{\tt
  hep-th/0005209}}].

\bibitem{Tanaka:2000jw}
T.~Tanaka, \emph{{A Comment on transPlanckian physics in inflationary
  universe}},  \href{http://arxiv.org/abs/astro-ph/0012431}{{\tt
  astro-ph/0012431}}.

\bibitem{Starobinsky:2001kn}
A.~A. Starobinsky, \emph{{Robustness of the inflationary perturbation spectrum
  to transPlanckian physics}},
  \href{http://dx.doi.org/10.1134/1.1381588}{\emph{Pisma Zh. Eksp. Teor. Fiz.}
  {\bf 73} (2001) 415--418}, [\href{http://arxiv.org/abs/astro-ph/0104043}{{\tt
  astro-ph/0104043}}].

\bibitem{Starobinsky:2002rp}
A.~A. Starobinsky and I.~I. Tkachev, \emph{{Trans-Planckian particle creation
  in cosmology and ultra-high energy cosmic rays}},
  \href{http://dx.doi.org/10.1134/1.1520612}{\emph{JETP Lett.} {\bf 76} (2002)
  235--239}, [\href{http://arxiv.org/abs/astro-ph/0207572}{{\tt
  astro-ph/0207572}}].

\bibitem{Hui:2001ce}
L.~Hui and W.~H. Kinney, \emph{{Short distance physics and the consistency
  relation for scalar and tensor fluctuations in the inflationary universe}},
  \href{http://dx.doi.org/10.1103/PhysRevD.65.103507}{\emph{Phys. Rev.} {\bf
  D65} (2002) 103507}, [\href{http://arxiv.org/abs/astro-ph/0109107}{{\tt
  astro-ph/0109107}}].

\bibitem{Niemeyer:2001qe}
J.~C. Niemeyer and R.~Parentani, \emph{{Transplanckian dispersion and scale
  invariance of inflationary perturbations}},
  \href{http://dx.doi.org/10.1103/PhysRevD.64.101301}{\emph{Phys. Rev.} {\bf
  D64} (2001) 101301}, [\href{http://arxiv.org/abs/astro-ph/0101451}{{\tt
  astro-ph/0101451}}].

\bibitem{Kaloper:2002uj}
N.~Kaloper, M.~Kleban, A.~E. Lawrence and S.~Shenker, \emph{{Signatures of
  short distance physics in the cosmic microwave background}},
  \href{http://dx.doi.org/10.1103/PhysRevD.66.123510}{\emph{Phys. Rev.} {\bf
  D66} (2002) 123510}, [\href{http://arxiv.org/abs/hep-th/0201158}{{\tt
  hep-th/0201158}}].

\bibitem{Kaloper:2002cs}
N.~Kaloper, M.~Kleban, A.~Lawrence, S.~Shenker and L.~Susskind, \emph{{Initial
  conditions for inflation}},
  \href{http://dx.doi.org/10.1088/1126-6708/2002/11/037}{\emph{JHEP} {\bf 11}
  (2002) 037}, [\href{http://arxiv.org/abs/hep-th/0209231}{{\tt
  hep-th/0209231}}].

\bibitem{Burgess:2002ub}
C.~P. Burgess, J.~M. Cline, F.~Lemieux and R.~Holman, \emph{{Are inflationary
  predictions sensitive to very high-energy physics?}},
  \href{http://dx.doi.org/10.1088/1126-6708/2003/02/048}{\emph{JHEP} {\bf 02}
  (2003) 048}, [\href{http://arxiv.org/abs/hep-th/0210233}{{\tt
  hep-th/0210233}}].

\bibitem{Brandenberger:2002hs}
R.~H. Brandenberger and J.~Martin, \emph{{On signatures of short distance
  physics in the cosmic microwave background}},
  \href{http://dx.doi.org/10.1142/S0217751X02010765}{\emph{Int. J. Mod. Phys.}
  {\bf A17} (2002) 3663--3680},
  [\href{http://arxiv.org/abs/hep-th/0202142}{{\tt hep-th/0202142}}].

\bibitem{Elgaroy:2003gq}
O.~Elgaroy and S.~Hannestad, \emph{{Can Planck-scale physics be seen in the
  cosmic microwave background?}},
  \href{http://dx.doi.org/10.1103/PhysRevD.68.123513}{\emph{Phys. Rev.} {\bf
  D68} (2003) 123513}, [\href{http://arxiv.org/abs/astro-ph/0307011}{{\tt
  astro-ph/0307011}}].

\bibitem{Greene:2004np}
B.~R. Greene, K.~Schalm, G.~Shiu and J.~P. van~der Schaar, \emph{{Decoupling in
  an expanding universe: Backreaction barely constrains short distance effects
  in the CMB}},
  \href{http://dx.doi.org/10.1088/1475-7516/2005/02/001}{\emph{JCAP} {\bf 0502}
  (2005) 001}, [\href{http://arxiv.org/abs/hep-th/0411217}{{\tt
  hep-th/0411217}}].

\bibitem{Greene:2005wk}
B.~Greene, M.~Parikh and J.~P. van~der Schaar, \emph{{Universal correction to
  the inflationary vacuum}},
  \href{http://dx.doi.org/10.1088/1126-6708/2006/04/057}{\emph{JHEP} {\bf 04}
  (2006) 057}, [\href{http://arxiv.org/abs/hep-th/0512243}{{\tt
  hep-th/0512243}}].

\bibitem{Danielsson:2002kx}
U.~H. Danielsson, \emph{{A Note on inflation and transPlanckian physics}},
  \href{http://dx.doi.org/10.1103/PhysRevD.66.023511}{\emph{Phys. Rev.} {\bf
  D66} (2002) 023511}, [\href{http://arxiv.org/abs/hep-th/0203198}{{\tt
  hep-th/0203198}}].

\bibitem{Danielsson:2002mb}
U.~H. Danielsson, \emph{{On the consistency of de Sitter vacua}},
  \href{http://dx.doi.org/10.1088/1126-6708/2002/12/025}{\emph{JHEP} {\bf 12}
  (2002) 025}, [\href{http://arxiv.org/abs/hep-th/0210058}{{\tt
  hep-th/0210058}}].

\bibitem{Danielsson:2002qh}
U.~H. Danielsson, \emph{{Inflation, holography, and the choice of vacuum in de
  Sitter space}},
  \href{http://dx.doi.org/10.1088/1126-6708/2002/07/040}{\emph{JHEP} {\bf 07}
  (2002) 040}, [\href{http://arxiv.org/abs/hep-th/0205227}{{\tt
  hep-th/0205227}}].

\bibitem{Danielsson:2005cc}
U.~H. Danielsson, \emph{{Inflation as a probe of new physics}},
  \href{http://dx.doi.org/10.1088/1475-7516/2006/03/014}{\emph{JCAP} {\bf 0603}
  (2006) 014}, [\href{http://arxiv.org/abs/hep-th/0511273}{{\tt
  hep-th/0511273}}].

\bibitem{Danielsson:2006gg}
U.~H. Danielsson, \emph{{Transplanckian signatures in WMAP3?}},
  \href{http://arxiv.org/abs/astro-ph/0606474}{{\tt astro-ph/0606474}}.

\bibitem{Goldstein:2002fc}
K.~Goldstein and D.~A. Lowe, \emph{{Initial state effects on the cosmic
  microwave background and transPlanckian physics}},
  \href{http://dx.doi.org/10.1103/PhysRevD.67.063502}{\emph{Phys. Rev.} {\bf
  D67} (2003) 063502}, [\href{http://arxiv.org/abs/hep-th/0208167}{{\tt
  hep-th/0208167}}].

\bibitem{Easther:2002xe}
R.~Easther, B.~R. Greene, W.~H. Kinney and G.~Shiu, \emph{{A Generic estimate
  of transPlanckian modifications to the primordial power spectrum in
  inflation}}, \href{http://dx.doi.org/10.1103/PhysRevD.66.023518}{\emph{Phys.
  Rev.} {\bf D66} (2002) 023518},
  [\href{http://arxiv.org/abs/hep-th/0204129}{{\tt hep-th/0204129}}].

\bibitem{Chung:2003wn}
D.~J.~H. Chung, A.~Notari and A.~Riotto, \emph{{Minimal theoretical
  uncertainties in inflationary predictions}},
  \href{http://dx.doi.org/10.1088/1475-7516/2003/10/012}{\emph{JCAP} {\bf 0310}
  (2003) 012}, [\href{http://arxiv.org/abs/hep-ph/0305074}{{\tt
  hep-ph/0305074}}].

\bibitem{Kaloper:2003nv}
N.~Kaloper and M.~Kaplinghat, \emph{{Primeval corrections to the CMB
  anisotropies}},
  \href{http://dx.doi.org/10.1103/PhysRevD.68.123522}{\emph{Phys. Rev.} {\bf
  D68} (2003) 123522}, [\href{http://arxiv.org/abs/hep-th/0307016}{{\tt
  hep-th/0307016}}].

\bibitem{Burgess:2003hw}
C.~P. Burgess, J.~M. Cline, F.~Lemieux and R.~Holman, \emph{{Decoupling,
  trans-planckia and inflation}},  in \emph{{Cosmic inflation. Proceedings,
  Meeting, Davis, USA, March 22-25, 2003}}, 2003.
\newblock \href{http://arxiv.org/abs/astro-ph/0306236}{{\tt astro-ph/0306236}}.

\bibitem{Alberghi:2003am}
G.~L. Alberghi, R.~Casadio and A.~Tronconi, \emph{{TransPlanckian footprints in
  inflationary cosmology}},
  \href{http://dx.doi.org/10.1016/j.physletb.2003.11.004}{\emph{Phys. Lett.}
  {\bf B579} (2004) 1--5}, [\href{http://arxiv.org/abs/gr-qc/0303035}{{\tt
  gr-qc/0303035}}].

\bibitem{Martin:2003kp}
J.~Martin and R.~Brandenberger, \emph{{On the dependence of the spectra of
  fluctuations in inflationary cosmology on transPlanckian physics}},
  \href{http://dx.doi.org/10.1103/PhysRevD.68.063513}{\emph{Phys. Rev.} {\bf
  D68} (2003) 063513}, [\href{http://arxiv.org/abs/hep-th/0305161}{{\tt
  hep-th/0305161}}].

\bibitem{Meerburg:2010rp}
P.~D. Meerburg and J.~P. van~der Schaar, \emph{{Minimal cut-off vacuum state
  constraints from CMB bispectrum statistics}},
  \href{http://dx.doi.org/10.1103/PhysRevD.83.043520}{\emph{Phys. Rev.} {\bf
  D83} (2011) 043520}, [\href{http://arxiv.org/abs/1009.5660}{{\tt
  1009.5660}}].

\bibitem{Kundu:2011sg}
S.~Kundu, \emph{{Inflation with General Initial Conditions for Scalar
  Perturbations}},
  \href{http://dx.doi.org/10.1088/1475-7516/2012/02/005}{\emph{JCAP} {\bf 1202}
  (2012) 005}, [\href{http://arxiv.org/abs/1110.4688}{{\tt 1110.4688}}].

\bibitem{Groeneboom:2007rf}
N.~E. Groeneboom and O.~Elgaroy, \emph{{Detection of transplanckian effects in
  the cosmic microwave background}},
  \href{http://dx.doi.org/10.1103/PhysRevD.77.043522}{\emph{Phys. Rev.} {\bf
  D77} (2008) 043522}, [\href{http://arxiv.org/abs/0711.1793}{{\tt
  0711.1793}}].

\bibitem{Ashoorioon:2013eia}
A.~Ashoorioon, K.~Dimopoulos, M.~M. Sheikh-Jabbari and G.~Shiu,
  \emph{{Reconciliation of High Energy Scale Models of Inflation with Planck}},
  \href{http://dx.doi.org/10.1088/1475-7516/2014/02/025}{\emph{JCAP} {\bf 1402}
  (2014) 025}, [\href{http://arxiv.org/abs/1306.4914}{{\tt 1306.4914}}].

\bibitem{Ashoorioon:2014nta}
A.~Ashoorioon, K.~Dimopoulos, M.~M. Sheikh-Jabbari and G.~Shiu,
  \emph{{Non-Bunch–Davis initial state reconciles chaotic models with BICEP
  and Planck}},
  \href{http://dx.doi.org/10.1016/j.physletb.2014.08.038}{\emph{Phys. Lett.}
  {\bf B737} (2014) 98--102}, [\href{http://arxiv.org/abs/1403.6099}{{\tt
  1403.6099}}].

\bibitem{Ashoorioon:2017toq}
A.~Ashoorioon, R.~Casadio, G.~Geshnizjani and H.~J. Kim, \emph{{Getting
  Super-Excited with Modified Dispersion Relations}},
  \href{http://dx.doi.org/10.1088/1475-7516/2017/09/008}{\emph{JCAP} {\bf 1709}
  (2017) 008}, [\href{http://arxiv.org/abs/1702.06101}{{\tt 1702.06101}}].

\bibitem{Broy:2016zik}
B.~J. Broy, \emph{{Corrections to $n_s$ and $n_t$ from high scale physics}},
  \href{http://dx.doi.org/10.1103/PhysRevD.94.109901,
  10.1103/PhysRevD.94.103508}{\emph{Phys. Rev.} {\bf D94} (2016) 103508},
  [\href{http://arxiv.org/abs/1609.03570}{{\tt 1609.03570}}].

\bibitem{Allen:1985ux}
B.~Allen, \emph{{Vacuum States in de Sitter Space}},
  \href{http://dx.doi.org/10.1103/PhysRevD.32.3136}{\emph{Phys. Rev.} {\bf D32}
  (1985) 3136}.

\bibitem{Mottola:1984ar}
E.~Mottola, \emph{{Particle Creation in de Sitter Space}},
  \href{http://dx.doi.org/10.1103/PhysRevD.31.754}{\emph{Phys. Rev.} {\bf D31}
  (1985) 754}.

\bibitem{DeWitt:2007mi}
B.~S. DeWitt and G.~Esposito, \emph{{An Introduction to quantum gravity}},
  \href{http://dx.doi.org/10.1142/S0219887808002679}{\emph{Int. J. Geom. Meth.
  Mod. Phys.} {\bf 5} (2008) 101--156},
  [\href{http://arxiv.org/abs/0711.2445}{{\tt 0711.2445}}].

\bibitem{Smolin:2003rk}
L.~Smolin, \emph{{How far are we from the quantum theory of gravity?}},
  \href{http://arxiv.org/abs/hep-th/0303185}{{\tt hep-th/0303185}}.

\bibitem{Giddings:2011dr}
S.~B. Giddings, \emph{{Is string theory a theory of quantum gravity?}},
  \href{http://dx.doi.org/10.1007/s10701-011-9612-x}{\emph{Found. Phys.} {\bf
  43} (2013) 115}, [\href{http://arxiv.org/abs/1105.6359}{{\tt 1105.6359}}].

\bibitem{birrell1984quantum}
N.~D. Birrell and P.~C.~W. Davies, \emph{Quantum fields in curved space}.
\newblock No.~7. Cambridge university press, 1984.

\bibitem{fulling1989aspects}
S.~A. Fulling, \emph{Aspects of quantum field theory in curved spacetime},
  vol.~17.
\newblock Cambridge university press, 1989.

\bibitem{parker2009quantum}
L.~Parker and D.~Toms, \emph{Quantum field theory in curved spacetime:
  quantized fields and gravity}.
\newblock Cambridge university press, 2009.

\bibitem{DeWitt:1975ys}
B.~S. DeWitt, \emph{{Quantum Field Theory in Curved Space-Time}},
  \href{http://dx.doi.org/10.1016/0370-1573(75)90051-4}{\emph{Phys. Rept.} {\bf
  19} (1975) 295--357}.

\bibitem{Bunch:1979uk}
T.~S. Bunch and L.~Parker, \emph{{Feynman Propagator in Curved Space-Time: A
  Momentum Space Representation}},
  \href{http://dx.doi.org/10.1103/PhysRevD.20.2499}{\emph{Phys. Rev.} {\bf D20}
  (1979) 2499--2510}.

\bibitem{buchbinder1980effective}
I.~Buchbinder, S.~Odintsov and I.~Shapiro, \emph{Effective action in quantum
  gravity (iop, bristol, 1992)}, {\emph{Google Scholar} (1980) 413}.

\bibitem{Shapiro:2008sf}
I.~L. Shapiro, \emph{{Effective Action of Vacuum: Semiclassical Approach}},
  \href{http://dx.doi.org/10.1088/0264-9381/25/10/103001}{\emph{Class. Quant.
  Grav.} {\bf 25} (2008) 103001}, [\href{http://arxiv.org/abs/0801.0216}{{\tt
  0801.0216}}].

\bibitem{Zeldovich:1971mw}
{\relax Ya}.~B. Zeldovich and A.~A. Starobinsky, \emph{{Particle production and
  vacuum polarization in an anisotropic gravitational field}}, {\emph{Sov.
  Phys. JETP} {\bf 34} (1972) 1159--1166}.

\bibitem{bunch1980adiabatic}
T.~Bunch, \emph{Adiabatic regularisation for scalar fields with arbitrary
  coupling to the scalar curvature}, {\emph{Journal of Physics A: Mathematical
  and General} {\bf 13} (1980) 1297}.

\bibitem{Parker:1974qw}
L.~Parker and S.~A. Fulling, \emph{{Adiabatic regularization of the energy
  momentum tensor of a quantized field in homogeneous spaces}},
  \href{http://dx.doi.org/10.1103/PhysRevD.9.341}{\emph{Phys. Rev.} {\bf D9}
  (1974) 341--354}.

\bibitem{Fulling:1974pu}
S.~A. Fulling, L.~Parker and B.~L. Hu, \emph{{Conformal energy-momentum tensor
  in curved spacetime: Adiabatic regularization and renormalization}},
  \href{http://dx.doi.org/10.1103/PhysRevD.10.3905}{\emph{Phys. Rev.} {\bf D10}
  (1974) 3905--3924}.

\bibitem{Fulling:1974zr}
S.~A. Fulling and L.~Parker, \emph{{Renormalization in the theory of a
  quantized scalar field interacting with a robertson-walker spacetime}},
  \href{http://dx.doi.org/10.1016/0003-4916(74)90451-5}{\emph{Annals Phys.}
  {\bf 87} (1974) 176--204}.

\bibitem{birrell1978application}
N.~Birrell, \emph{The application of adiabatic regularization to calculations
  of cosmological interest},  in \emph{Proceedings of the Royal Society of
  London A: Mathematical, Physical and Engineering Sciences}, vol.~361,
  pp.~513--526, The Royal Society, 1978.

\bibitem{Anderson:1987yt}
P.~R. Anderson and L.~Parker, \emph{{Adiabatic Regularization in Closed
  Robertson-walker Universes}},
  \href{http://dx.doi.org/10.1103/PhysRevD.36.2963}{\emph{Phys. Rev.} {\bf D36}
  (1987) 2963}.

\bibitem{Haro:2010zz}
J.~Haro, \emph{{Calculation of the renormalized two-point function by adiabatic
  regularization}},
  \href{http://dx.doi.org/10.1007/s11232-010-0123-2}{\emph{Theor. Math. Phys.}
  {\bf 165} (2010) 1490--1499}.

\bibitem{Haro:2010mx}
J.~Haro, \emph{{Topics in Quantum Field Theory in Curved Space}},
  \href{http://arxiv.org/abs/1011.4772}{{\tt 1011.4772}}.

\bibitem{Kohri:2017iyl}
K.~Kohri and H.~Matsui, \emph{{Electroweak Vacuum Instability and Renormalized
  Vacuum Field Fluctuations in Friedmann-Lemaitre-Robertson-Walker
  Background}}, \href{http://dx.doi.org/10.1103/PhysRevD.98.103521}{\emph{Phys.
  Rev.} {\bf D98} (2018) 103521}, [\href{http://arxiv.org/abs/1704.06884}{{\tt
  1704.06884}}].

\bibitem{Ringwald:1987ui}
A.~Ringwald, \emph{{Evolution Equation for the Expectation Value of a Scalar
  Field in Spatially Flat Rw Universes}},
  \href{http://dx.doi.org/10.1016/S0003-4916(87)80027-1}{\emph{Annals Phys.}
  {\bf 177} (1987) 129}.

\bibitem{Riotto:2002yw}
A.~Riotto, \emph{{Inflation and the theory of cosmological perturbations}},
  {\emph{ICTP Lect. Notes Ser.} {\bf 14} (2003) 317--413},
  [\href{http://arxiv.org/abs/hep-ph/0210162}{{\tt hep-ph/0210162}}].

\bibitem{Finelli:2008zg}
F.~Finelli, G.~Marozzi, A.~A. Starobinsky, G.~P. Vacca and G.~Venturi,
  \emph{{Generation of fluctuations during inflation: Comparison of stochastic
  and field-theoretic approaches}},
  \href{http://dx.doi.org/10.1103/PhysRevD.79.044007}{\emph{Phys. Rev.} {\bf
  D79} (2009) 044007}, [\href{http://arxiv.org/abs/0808.1786}{{\tt
  0808.1786}}].

\bibitem{Finelli:2010sh}
F.~Finelli, G.~Marozzi, A.~A. Starobinsky, G.~P. Vacca and G.~Venturi,
  \emph{{Stochastic growth of quantum fluctuations during slow-roll
  inflation}}, \href{http://dx.doi.org/10.1103/PhysRevD.82.064020}{\emph{Phys.
  Rev.} {\bf D82} (2010) 064020}, [\href{http://arxiv.org/abs/1003.1327}{{\tt
  1003.1327}}].

\bibitem{Boyanovsky:2005sh}
D.~Boyanovsky, H.~J. de~Vega and N.~G. Sanchez, \emph{{Quantum corrections to
  slow roll inflation and new scaling of superhorizon fluctuations}},
  \href{http://dx.doi.org/10.1016/j.nuclphysb.2006.04.010}{\emph{Nucl. Phys.}
  {\bf B747} (2006) 25--54}, [\href{http://arxiv.org/abs/astro-ph/0503669}{{\tt
  astro-ph/0503669}}].

\bibitem{Sloth:2006az}
M.~S. Sloth, \emph{{On the one loop corrections to inflation and the CMB
  anisotropies}},
  \href{http://dx.doi.org/10.1016/j.nuclphysb.2006.04.029}{\emph{Nucl. Phys.}
  {\bf B748} (2006) 149--169},
  [\href{http://arxiv.org/abs/astro-ph/0604488}{{\tt astro-ph/0604488}}].

\bibitem{Linde:1982uu}
A.~D. Linde, \emph{{Scalar Field Fluctuations in Expanding Universe and the New
  Inflationary Universe Scenario}},
  \href{http://dx.doi.org/10.1016/0370-2693(82)90293-3}{\emph{Phys. Lett.} {\bf
  116B} (1982) 335--339}.

\bibitem{Vilenkin:1983xp}
A.~Vilenkin, \emph{{Quantum Fluctuations in the New Inflationary Universe}},
  \href{http://dx.doi.org/10.1016/0550-3213(83)90208-0}{\emph{Nucl. Phys.} {\bf
  B226} (1983) 527--546}.

\bibitem{Ade:2015lrj}
{\scshape Planck} collaboration, P.~A.~R. Ade et~al., \emph{{Planck 2015
  results. XX. Constraints on inflation}},
  \href{http://dx.doi.org/10.1051/0004-6361/201525898}{\emph{Astron.
  Astrophys.} {\bf 594} (2016) A20},
  [\href{http://arxiv.org/abs/1502.02114}{{\tt 1502.02114}}].

\bibitem{Ade:2015tva}
{\scshape BICEP2, Planck} collaboration, P.~A.~R. Ade et~al., \emph{{Joint
  Analysis of BICEP2/$Keck Array$ and $Planck$ Data}},
  \href{http://dx.doi.org/10.1103/PhysRevLett.114.101301}{\emph{Phys. Rev.
  Lett.} {\bf 114} (2015) 101301}, [\href{http://arxiv.org/abs/1502.00612}{{\tt
  1502.00612}}].

\bibitem{Bezrukov:2007ep}
F.~L. Bezrukov and M.~Shaposhnikov, \emph{{The Standard Model Higgs boson as
  the inflaton}},
  \href{http://dx.doi.org/10.1016/j.physletb.2007.11.072}{\emph{Phys. Lett.}
  {\bf B659} (2008) 703--706}, [\href{http://arxiv.org/abs/0710.3755}{{\tt
  0710.3755}}].

\bibitem{Barvinsky:2008ia}
A.~O. Barvinsky, A.~{\relax Yu}. Kamenshchik and A.~A. Starobinsky,
  \emph{{Inflation scenario via the Standard Model Higgs boson and LHC}},
  \href{http://dx.doi.org/10.1088/1475-7516/2008/11/021}{\emph{JCAP} {\bf 0811}
  (2008) 021}, [\href{http://arxiv.org/abs/0809.2104}{{\tt 0809.2104}}].

\bibitem{Salvio:2015kka}
A.~Salvio and A.~Mazumdar, \emph{{Classical and Quantum Initial Conditions for
  Higgs Inflation}},
  \href{http://dx.doi.org/10.1016/j.physletb.2015.09.020}{\emph{Phys. Lett.}
  {\bf B750} (2015) 194--200}, [\href{http://arxiv.org/abs/1506.07520}{{\tt
  1506.07520}}].

\bibitem{Calmet:2016fsr}
X.~Calmet and I.~Kuntz, \emph{{Higgs Starobinsky Inflation}},
  \href{http://dx.doi.org/10.1140/epjc/s10052-016-4136-3}{\emph{Eur. Phys. J.}
  {\bf C76} (2016) 289}, [\href{http://arxiv.org/abs/1605.02236}{{\tt
  1605.02236}}].

\bibitem{Martin:2013tda}
J.~Martin, C.~Ringeval and V.~Vennin, \emph{{Encyclopædia Inflationaris}},
  \href{http://dx.doi.org/10.1016/j.dark.2014.01.003}{\emph{Phys. Dark Univ.}
  {\bf 5-6} (2014) 75--235}, [\href{http://arxiv.org/abs/1303.3787}{{\tt
  1303.3787}}].

\bibitem{Martin:2013nzq}
J.~Martin, C.~Ringeval, R.~Trotta and V.~Vennin, \emph{{The Best Inflationary
  Models After Planck}},
  \href{http://dx.doi.org/10.1088/1475-7516/2014/03/039}{\emph{JCAP} {\bf 1403}
  (2014) 039}, [\href{http://arxiv.org/abs/1312.3529}{{\tt 1312.3529}}].

\bibitem{Martin:2013gra}
J.~Martin, \emph{{Inflation after Planck: and the winners are}},  in
  \emph{{Rencontres du Vietnam: Hot Topics in General Relativity and
  Gravitation (HTGRG-1) Quy Nhon, Vietnam, 28 July—3 August 2013}}, 2013.
\newblock \href{http://arxiv.org/abs/1312.3720}{{\tt 1312.3720}}.

\bibitem{Ijjas:2013vea}
A.~Ijjas, P.~J. Steinhardt and A.~Loeb, \emph{{Inflationary paradigm in trouble
  after Planck2013}},
  \href{http://dx.doi.org/10.1016/j.physletb.2013.05.023}{\emph{Phys. Lett.}
  {\bf B723} (2013) 261--266}, [\href{http://arxiv.org/abs/1304.2785}{{\tt
  1304.2785}}].

\end{thebibliography}\endgroup
\bibliographystyle{JHEP}
%%%%%%%%%%%%%%%%%%%%%%%%%%%%

\end{document}